\documentclass[preprints,article,accept,moreauthors,pdftex]{Definitions/mdpi} 
\usepackage{graphicx}
\usepackage{enumitem}

\firstpage{1} 
\pubvolume{1}
\issuenum{1}
\articlenumber{0}
\pubyear{2021}
\copyrightyear{2020}
\datereceived{} 
\dateaccepted{} 
\datepublished{} 
\hreflink{https://doi.org/} 

\Title{Estimating the parameters of Hybrid Palatini gravity model with the Schwarzschild precession of S2, S38 and S55 stars: case of bulk mass distribution}

\TitleCitation{Estimating the parameters of Hybrid Palatini gravity model with the Schwarzschild precession of S2, S38 and S55 stars: case of bulk mass distribution}

\Author{Du\v{s}ko Borka$^{1,*}$\orcidA{}, Vesna Borka Jovanovi\'{c}$^{1}$\orcidB{}, Violeta N. Nikoli\'{c}$^{1}$\orcidC{}, Nenad Dj. Lazarov$^{1}$ and Predrag Jovanovi\'{c}$^{2}$\orcidE{}}

\AuthorNames{Du\v{s}ko Borka, Vesna Borka Jovanovi\'{c}, Violeta N. Nikoli\'{c}, Nenad Dj. Lazarov and Predrag Jovanovi\'{c}}

\AuthorCitation{Borka, D.; Borka Jovanovi\'{c}, V., Nikoli\'{c}, V. N.; Lazarov, N. Dj.; Jovanovi\'{c}, P.}

\address{%
$^{1}$ \quad Department of Theoretical Physics and Condensed Matter Physics (020), Vin\v{c}a Institute of Nuclear Sciences - National Institute of the Republic of Serbia, University of Belgrade, P.O. Box 522, 11001 Belgrade, Serbia \\
$^{2}$ \quad Astronomical Observatory, Volgina 7, P.O. Box 74, 11060 Belgrade, Serbia}

\corres{Correspondence: dusborka@vinca.rs}

\abstract{We estimate the parameters of Hybrid Palatini gravity model with the Schwarzschild precession of S-stars, specifically of S2, S38 and S55 stars. We also take into account case of bulk mass distribution near Galactic Center. We assume that the Schwarzschild orbital precession of mentioned S-stars is the same like in General Relativity (GR) in all studied cases. In 2020 the GRAVITY Collaboration detected the orbital precession of the S2 star around the supermassive black hole (SMBH) at the Galactic Center and showed that it is close to the GR prediction. The astronomical data analysis of S38 and S55 orbits showed that also in these cases orbital precession is close to the GR prediction. Based on this observational fact, we evaluated parameters of the Hybrid Palatini Gravity model with the Schwarzschild precession of the S2, S38 and S55 stars and we estimated the range of parameters of Hybrid Palatini gravity model for which the orbital precession is like in GR for all tree stars. We also evaluated parameters of the Hybrid Palatini Gravity model in case of different values of bulk mass density distribution of extended matter. We believe that proposed method is a useful tool to evaluate parameters of the gravitational potential at the Galactic Center.}


\begin{document}

\section{Introduction}

In recent few decades, various modified gravity theories have appeared as potential extensions of Einstein's gravity theory \cite{fisc99}. One of the reasons for a postulation of mentioned theories is the possibility to exclude concept of dark energy and dark matter, and to explain  cosmological and astrophysical data collected at different scales just considering further degrees of freedom of the gravitational field. This occurs as a consequence of geometric corrections \cite{capo11b}. Modified gravity theories have to resolve different observations concerning, starting from the Solar system, neutron stars, binary pulsars, spiral and elliptical galaxies, clusters of galaxies, up to the large-scale structure of the Universe  \cite{noji11,noji17,capo12a,salu21,dimi19}. That is, in Ref. \cite{noji11} a cosmological reconstruction (characterized by a very general character) of various modified gravity is given, and in \cite{noji17} various formalisms of representatives $(F(R), F(G), F(T))$ of standard modified gravity are presented, as well as alternative theoretical approaches. Ref. \cite{capo12a} described stars and cluster of galaxies (spiral and elliptical galaxies, also), beyond the scope of dark matter, by extending the Hilbert-Einstein action to $f(R)$ gravity, and in \cite{salu21} the authors discussed observations and experiments, which depicted to the fact that GR and the Standard model of elementary particles are unable to explain the phenomena behind the dark matter concept. In \cite{dimi19}, the chosen cosmological parameters are determined (as accurate cosmological solutions) within the framework of represented nonlocal gravitational model, which showed satisfactorily agreement with experimental observations.

\subsection*{Some alternative theories of gravity}

Let us recall that numerous alternative gravity theories have been proposed (see e.g. \cite{clif06,kope04,ruba08,babi10,capo11a,clif12,rham14,mart15,rham17,petr17,bork21b}). For example, the alternative theories of gravity are discussed in \cite{clif06}. In \cite{kope04}, the authors introduced extension of the post-Newtonian relativistic theory by considering additionally all relativistic effects, which originated from the presumable existence of a scalar field. Ref. \cite{ruba08} presents a review article, in which the authors discussed chosen aspects of 4D massive gravities. In Ref. \cite{babi10} a numerical solution of the nonlinear Pauli--Fierz theory is given. The proposed solution represents an improvement of the existing solution of GR, which was achieved by including the Vainshtein mechanism. In \cite{capo11a}, the extended theories of gravity were discussed, by taking into account $f(R)$ and scalar-tensor gravity in metric and Palatini approaches; the issues like inflation, large scale structure, dark energy, dark matter, quantum gravity, were discussed also. Ref. \cite{clif12} represents review of modified theories of gravity and models of extra dimensions, such as: Scalar-Tensor, Einstein-Aether, Bimetric theories, TeVeS, $F(R)$, Horava-Lifschitz gravity, Galileons, Ghost Condensates, Kaluza-Klein, Randall-Sundrum, DGP, higher co-dimension braneworlds, as well as construction of Parametrized Post-Friedmannian formalism. In paper \cite{rham14}, the Dvali-Gabadadze-Porrati model (DGP), cascading gravity, ghost-free massive gravity, new mass gravity, Lorentz-violating massive gravity and non-local massive gravity are discussed. The $f(R)$ modifications of general relativity, considering galaxy clusters, cosmological perturbations, and N-body simulations, are discussed in \cite{mart15}. A few observational mass bounds have been established, and among them, the mass bounds from the effects of the Yukawa potential in Ref. \cite{rham17}. Ref. \cite{petr17} presents monograph in which the mathematical background is given (for example, conservation laws and symmetries for different theories of gravity), necessary for comparison of methods of perturbations in general relativity; this mathematical introduction enables building of different modified-gravity theories. In paper \cite{bork21b}, it is given the method for evaluation of parameters of the gravitational potential at the Galactic Center, based on the extended gravity models (power-law $f(R)$, general Yukawa-like corrections, scalar-tensor gravity, and non-local gravity theories formulated in both metric and Palatini formalism).

\subsection*{Some alternative approaches for the weak field limit of theories of gravity}

Noteworthy, different alternative approaches for the weak field limit (starting from fourth-order theories of gravity, like $f(R)$), have been proposed and considered \cite{zakh06,zakh07,frig07,nuci07,zakh09,capo09a,iori07,iori08,iori10,bork12,capo14,doku15,doku17,doku15b}. For example, in Ref. \cite{zakh06} the gravitational microlensing is discussed, considered from the aspect of the weak field limit of fourth order gravity theory, and in \cite{zakh07} determination of the mass and the size of dark matter sphere is discussed, based on the $\gamma$-ray emission from the Galactic Center region. Ref. \cite{frig07} examined the consequences of modified $f(R)$ gravity (power-law $f(R)$) on galactic scales, by performing an analysis of rotational curves. In Ref. \cite{nuci07} it is discussed the search for general relativistic periastronic shifts, which is limited by the existence of clusters around black hole, which could modify orbits due to classical effects that mask the general relativistic effect. Ref. \cite{zakh09} represents a discussion of solving the problem of dark matter and dark energy (which could be done by considering changing the fundamental law of gravity). Ref. \cite{capo09a} shown that the metric approach of any analytic $f(R)$-gravity model, presents a weak field limit (the standard Newtonian potential is corrected by a Yukawa-like term), and \cite{iori07} considered the limitations of the range parameters $\lambda$, that are described by modifications of Newton's inverse square law of the gravity similar to Yukawa; the results of this study could affect all modified theories of gravity, which include Yukawa-type terms (which are characterized by a range of parameters much larger than the size of the solar system). In \cite{iori08} a Yukawa-like long-range modified model of gravity (MOG) is discussed. Ref. \cite{iori10} considered the Modified Newtonian Dynamics, introducing the integration of the equations of motion of Magellanic clouds in a numerical manner. In paper \cite{bork12}, the limitation of the $R^n$ gravity at Galactic scales, based on the simulation of the S2-like stars orbits, is discussed; it was shown that $R^n$ gravity impacts the simulated orbits in the qualitatively similar way as a bulk distribution of matter in Newton's gravity. In Ref. \cite{capo14}, an analytic fourth-order theory of gravity (which is non-minimally coupled with a massive scalar field) is applied, to explain deviations of S2 star orbit, by using gravitational potentials derived from modified gravity models in absence of dark matter. Refs. \cite{doku15,doku17} considered an analytical expression for the precession angle (with assumption of a power-law profile of the dark matter density); they calculated the mass of the dark matter in the vicinity of a supermassive black hole at the Galaxy center, based on the observations of nonrelativistic precession of the orbits of S0 stars. While in \cite{doku15b} the authors discussed the physical processes, occurred at the center of the Galaxy; results of this study revealed the mass of the SMBH Sgr A$^\ast$.

\subsection*{Some experimental limits related to Extended Theories of gravity}

Also, literature review revealed presence of some experimental limits related to Extended Theories of gravity \cite{avil12,duns16,capo13c,laur18a,mart18,laur18b,capo19,ana21,mart21}. In Ref. \cite{avil12}, the authors used cosmography to examine the kinematics of the Universe, by combination of theoretical derivation of cosmological distances and numerical data fitting, while in \cite{duns16} it is investigated whether cosmography could be used to ensure information on the cosmological expansion history and discussed the limits of experimentally probing of cosmographic expansion. In Ref. \cite{capo13c}, the authors performed cosmographic analyses and discussed the cosmological consequences of $f(R)$ and $f(T)$ gravities, as well as their influence on the cosmography framework. Also, they depicted to the unfavorable degeneracy problem (cosmographic constraints on $f(R)$ and $f(T)$ cannot be distinguished by theories of GR extensions and dark energy models). In \cite{laur18a} the differences between the Newtonian and relativistic approaches are described, and it is revealed that the relativistic approach presents more suitable strategy for further probing of modified theories of gravity. In Ref. \cite{mart18} the generalization of the gravitational action to a function $f(R)$ is investigated, as an alternative to the dark matter and dark energy, and also the weak field limit of the $f(R)$-gravity is discussed. In \cite{laur18b} the analytical $f(R)$-gravity model is considered, which is characterized by a Yukawa-like modification of the Newtonian potential, and leads to a modification of particle dynamics. In paper \cite{capo19} the authors performed a comparison between the $\Lambda CDM$ cosmological model and $f(R)$ and $f(T)$ models; they presented a new approach to breaking degeneration among dark energy models, introduced to overcome the limits of standard cosmography. The reference \cite{ana21} discussed on the usage of S-stars observations to constrain a Yukawa-like gravitational potential and considered the fact that deviations from GR are parametrized by the strength of the potential, $\delta$, and its length scale, $\lambda$. In \cite{mart21}, it is showed that the observing stars orbiting closer to the central gravitational source could allow to distinguish between the black hole and wormhole nature of this object (by observing S2 and S62 stars).

\subsection*{Gravitational potentials and the stellar dynamics}

In this study, gravitational potentials of self-gravitating structures have been investigated by considering the stellar dynamics. Recall that S-stars are the bright stars which move around the Galactic Center \cite{ghez00,scho02,gill09a,gill09b,ghez08,genz10,meye12,gill17,hees17,chu17,abut18,amor19,do19,abut19,hees20,abut20} where Sgr A$^\ast$ (which presents a compact massive object) is located. The conventional model, used to describe the Galactic Center, considers the SMBH with mass around $4.3\times 10^6 M_\odot$ and an extended mass distribution formed with stellar cluster and dark matter. Spherical shell, where trajectories of bright stars are located, should be characterized by a total mass of bulk distribution, which is significantly smaller compared to the black hole mass. In Ref. \cite{ghez00} measurements of the accelerations for three stars located $\sim 0.005$ pc from the central radio source Sgr A$^\ast$ are discussed; obtained data revealed the localization of the dark mass to within $0.05 \pm 0.04$ arcsec of the nominal position of Sgr A$^\ast$. Also, in \cite{ghez08} astrometric and radial velocity measurements, performed by the Keck telescopes, are discussed, as well as estimated distance ($R0$) and the Galaxy's local rotation speed; further, it was noticed increased black hole mass, depicted to the longer period for the innermost stable orbit, and longer resonant relaxation timescales for stars in the vicinity of the black hole. The authors of paper \cite{gill09a} discussed a moderate improvement of the statistical errors of mass and distance to Sgr A$^\ast$, and in \cite{gill09b} the orbits of 38 stars (among them, the orbit of the S2 star) are determined; all stellar orbits are fitted satisfactorily by a single point mass potential. In Ref. \cite{scho02}, the high resolution astrometric imaging is discussed, which is used to investigate two thirds of the orbit of the star currently closest to the massive black hole candidate SgrA$^\ast$; it is found that the star is on a bound, highly elliptical Keplerian orbit around SgrA$^\ast$, with an orbital period of 15.2 years and a pericentre distance of only 17 light hours. The authors in \cite{genz10} consider a massive black hole in the Galactic Center and a nuclear star cluster, by analyzing the size and motion measurements of the radio source Sgr A$^\ast$, which is understood as a massive black hole surrounded by a dense nuclear star cluster. In Ref. \cite{meye12}, the authors examined the behavior of a supermassive black hole by investigating stars with short orbital periods at the center of our galaxy; measurements from the Keck Observatory discovered the star S0-102 orbiting a black hole with a period of less than 15 years. Ref. \cite{gill17} represents an update of the main conclusions regarding the measurement of mass and distance to Sgr A$^\ast$, derived from data obtained by monitoring stellar orbits in the Galactic Center. In Ref. \cite{hees17} it is shown that short-period stars orbiting around the supermassive black hole in our Galactic Center can successfully be used to probe the gravitational theory in a strong regime. In \cite{chu17}, the behavior of the star S2, which orbits a supermassive black hole in a short period of time (less than 20 years) is considered; the authors reported on the first binarity limits of S0-2, observed from radial velocity monitoring. The GRAVITY Collaboration \cite{abut18} discussed the orbit of the S2 star around the massive black hole Sgr A$^\ast$, which is used as probe of the gravitational field in the center of the Galaxy; by using different statistical analysis methods, the authors detected the combined gravitational redshift and relativistic transverse Doppler effect for the S2 star, and found that the S2 data are not consistent with pure Newtonian dynamics. Also, in \cite{abut19} they presented the result of the measurement of the $R0$ (the geometric distance to the Galactic Center), by probing the S2 star, which is orbiting around the SMBH Sgr A$^\ast$. In Ref. \cite{do19}, the authors examined the prediction of GR (that a star passing near a SMBH shows a relativistic redshift), by using observations of the Galactic Center star S2; a combination of special relativistic- and gravitational-redshift was discovered, which confirms the model of General Relativity and excludes Newtonian's model. Ref. \cite{amor19} considered the assumption of the presence of a scalar field structure associated with a black hole at the center of our galaxy; the authors used the results of the orbital perturbation theory to compute the extent to which the orbital parameters of the S2 star change during the orbital period. Ref. \cite{hees20} introduced a new ways of probing fundamental physics, tracking stars in the Galactic Center; a new way of looking for changes in the fine structure constant has been proposed, by using measurements of late-type evolved giant stars from the S-star cluster orbiting a supermassive black hole in our Galactic Center. Ref. \cite{abut20} reported the first detection of the GR Schwarzschild precession in S2's orbit.

Ruffini, Arg\"uelles \& Rueda \cite{ruff15} discussed a dark matter distribution, and proposed that it is consisted of a dense core and a diluted halo. The dark matter distribution was named as the RAR-model. In 2021, Becerra-Vergara et al. \cite{bece21} commented this model, and concluded that the mentioned model ensures a better fit of bright stars trajectories, compared to the SMBH model. The properties of bright star trajectories in gravitational field of a dense core, described by RAR-model, have been discussed in \cite{zakh21}. In such case, trajectories of stars are ellipses like in Kepler's two-body problem but with one big difference: instead of their foci, the centers of the ellipses coincide with a Galactic Center and their orbital periods do not depend on their semi-mejor axes. Therefore, these properties are not consistent with existing observational data \cite{zakh21}. The orbital precession occurs as a consequence of relativistic effects, as well as of extended mass distribution, because both effects could cause perturbation of the Newtonian potential. In the first case, the precession induces a prograde pericentre shift, while in the second case retrograde shift occurs \cite{rubi01}. In both cases, as a final result, rosette shaped orbits are obtained \cite{adki07,wein05}. 

Besides Schwarzschild precession, relativistic frame-dragging due to the spin of SMBH, also known as the Lense-Thirring (LT) effect, could cause orbital precession. The LT precession in the case of several S-stars was studied in references \cite{peib20,iori20,iori21,gain20,bork21b}, and it was found that it is much smaller than Schwarzschild precession \cite{iori20,bork21b}. The spin of Sgr A$^\ast$ was estimated to $\chi_g < 0.1$ by the observed distribution of the orbital planes of the S-stars \cite{frag20}. Also, in this paper we considered only the solutions of Hybrid Palatini gravity model for spherically symmetric and stationary gravitational field, which do not include the SMBH spin. Having this in mind, we did not take into account the LT precession in our calculations for S-stars precession.

In our previous studies, we considered various Extended Gravity theories and compared theoretical models with astronomical data for different astrophysical scales: the S2 star orbit \cite{bork12,bork13,zakh14,bork16,zakh16,zakh18,zakh18a,dial19,bork19b,jova21}, fundamental plane of elliptical galaxies \cite{bork16a,capo20,bork21} and barionic Tully-Fischer relation of spiral galaxies \cite{capo17}. In this study, as a continuation of our previous paper \cite{bork21b}, the parameters of the Hybrid Palatini gravity model will be evaluated by Schwarzschild precession of S2, S38 and S55 stars. Here, we will also take into account the bulk mass density distribution of extended matter in the Galactic Center, and assume that the orbital precession of the S2, S38 and S55 stars are equal to the corresponding GR predictions of $0^\circ.18$, $0^\circ.11$ and $0^\circ.10$ per orbital period, respectively. We use this assumption because the GRAVITY Collaboration detected the orbital precession of the S2 star around the SMBH \cite{abut20} and showed that it is close to the corresponding prediction of GR. Also, according to data analysis in the framework of Yukawa gravity model in the paper \cite{ana21}, the orbital precessions of the S38 and S55 stars are close to the corresponding prediction of GR for these stars.

The paper is organized in the following way. In Section 2 we presented basics of Hybrid Palatini theoretical model, as well as model for bulk mass density distribution of extended matter. In Section 3 we evaluated parameters of Hybrid Palatini theoretical model by Schwarzschild precession of S2, S38 and S55 stars and discussed the obtained results. Concluding remarks are given in Section 4, while the Appendix A contains the detailed derivation of gravitational potential in the weak field limit for this gravity model.

\section{Theory}

In this article, we found constraints on parameters of Hybrid Palatini gravity model with request that the obtained values of orbital precession angles are the same like in GR, but for different values of mass density of matter. We used a weak field limit for Hybrid Palatini gravitation potential. A straightforward extension of GR is $f(R)$ gravity which, instead of the Einstein-Hilbert action (linear in the Ricci scalar $R$), considers a generic function of $R$ \cite{clif05,capo06,zakh06,zakh07,capo07,frig07,soti10}.

\subsection{Modified Hybrid Palatini gravity model}

There are two variational principles that one can apply to the Einstein-Hilbert action in order to derive Einstein’s equations: the standard metric variation and the Palatini variation \cite{soti10,olmo11,capo15}. The choice of the variational principle is usually referred to as a formalism, so one can use the terms metric or second-order formalism and Palatini or first-order formalism. In the Palatini variation the metric and the connection are assumed to be independent variables and one varies the action with respect to both of them. This variation leads to Einstein’s equations, under the important assumption that the matter action does not depend on the connection. Both variational principles lead to the same field equation for an action whose Lagrangian is linear in $R$, for example in the context of GR, but not for a more general action, for example in extended gravities. $f(R)$ gravity in the metric formalism is called metric $f(R)$ gravity, and $f(R)$ gravity in the Palatini formalism is called Palatini $f(R)$ gravity. The Palatini variational approach leads to second order differential field equations, while the resulting field equations in the metric approach are fourth order coupled differential equations \cite{soti10,olmo11,capo15}. There is also a novel approach, the hybrid variation of these theories. It consists of adding to the metric Einstein–Hilbert Lagrangian an  $f(R)$ term constructed within the framework of the Palatini formalism, i.e. purely metric Einstein-Hilbert action is supplemented with metric-affine correction terms constructed as Palatini \cite{hark12,capo13,capo13a,capo13b}. The $f(R)$ theories are the special limits of the one-parameter class of theories where the scalar field depends solely on the stress energy trace $T$ (Palatini version) or solely on the Ricci curvature $R$ (metric version). Here, we consider the hybrid metric-Palatini gravitational theory. In the general case, the field equations are fourth order both in the matter and in the metric derivatives. Hybrid metric-Palatini theory provides a unique interpolation between the two \textit{a priori} completely distinct classes of gravity theories. The aim of this formulation has twofold benefit: from one side, one wants to describe the extra gravitational budget in metric-affine formalism, from the other side, one wants to cure the shortcomings emerging in $f(R)$ gravity both in metric and Palatini formulations. In particular, hybrid gravity allows to disentangle the metric and the geodesic structures pointing out that further degrees of freedom coming from $f(R)$ can be recast as an auxiliary scalar field. An interesting aspect of metric-Palatini theories is the possibility to generate long-range forces without entering into conflict with local tests of gravity. Also, the possibility of expressing these hybrid $f(R)$ metric-Palatini theories using a scalar-tensor representation simplifies the analysis of the field equations and the construction of solutions. To get deeper insight, see  \cite{olmo11,hark12,capo13,capo13a,capo13b,koiv10,capo12}. 

The Palatini formalism and the metric one are completely different both from a qualitative and from a quantitative viewpoint. In the Palatini formalism field equations are easily solvable \cite{alle05}. In this sense the Palatini formalism is more easy to handle and simpler to analyze than the corresponding metric formalism. It is obvious that any reasonable model of gravity should satisfy the standard solar system tests. It has been shown that, in principle, Palatini formalism provides a good Newtonian approximation. Also, it is known that on-shell formulation of Palatini gravity coincides with that of same metric gravity \cite{alle05}. In paper \cite{boro21} a class of scalar-tensor theories is proposed including a non-metricity 	that unifies metric, Palatini and hybrid metric-Palatini gravitational actions with non-minimal interaction. Authors present a new approach to scalar-tensor theories of gravity that unifies: metric, Palatini, and hybrid. Such an approach will encompass within one family of theories not only metric, but also Palatini scalar-tensor theories of gravity, and will be a natural extension of the hybrid metric-Palatini gravity. It is shown that every such theory can be represented on-shell by a purely metric scalar-tensor theories possessing the same solutions for a metric and a scalar field.

Recall, in the weak field limit (see the Appendix for detailed explanation), the scalar field behaves as $\phi(r) \approx \phi_0 + \dfrac{2G\phi_0 M}{3rc^{2}} e^{-m_\phi r}$, where $M$ is the mass of the system and $r$ is the interaction length. Leading parameters for Hybrid Palatini gravity are $m_\phi$ and $\phi_0$. The aim of this study was to evaluate these parameters. We can write the modified gravitational potential in the following form \cite{capo13,bork16}:

\begin{equation}
\Phi \left( r \right) = -\dfrac{G}{1+\phi_0}\left[1-\left(\phi_0/3\right)e^{-m_\phi r}\right] M/r.
\label{equ01}
\end{equation}

The parameter $m_{\phi}$ represents a scaling parameter for gravity interaction and $[m_{\phi}]=[Length]^{-1}$. We measured the parameter in AU$^{-1}$ (AU is the astronomical unit). The parameter $\phi_0$ represents the amplitude of the background value of the scalar field $\phi$ and it is dimensionless. Non-zero values of these two parameters, if obtained, would indicate a potential deviation from GR.

\subsection{Orbital precession in case of bulk mass distribution}

In this study, we investigated S2, S38, and S55 stars. Orbital precession of investigated stars is influenced by other stars, gas, and dark matter. It is expected that the stars represent the dominant component of the extended galactic mass distribution near the central SMBH. To investigate orbital precession of S-stars, we made two assumptions. First we suppose presence of the Hybrid Palatini gravitational potential \cite{bork16}. Second assumption is a bulk distribution of mass around SMBH in the central regions of our Galaxy \cite{jova21}:

\begin{equation}
M(r)=M_{SMBH}+M_{ext}(r).
\label{equ02}
\end{equation}

A bulk mass distribution $M(r)$ consists of the central black hole of mass $M_{SMBH}= 4.3 \times10^6 M_\odot$ \cite{gill09a} and extended mass distribution $M_{ext}(r)$ enclosed within some radius $r$. $M_{ext}(r)$ is the total mass including the stellar cluster, interstellar gas and dark matter. To describe the mass density distribution of extended matter, we adopted a double power-law mass density profile \cite{genz03,pret09,amor19}:

\begin{equation}
\rho(r)=\rho_0\left( \dfrac{r}{r_0}\right) ^{-\alpha},\;\alpha =\left\{
\begin{array}{ll}
2.0\pm 0.1, & r\geq r_0 \\
1.4\pm 0.1, & r < r_0
\end{array}
\right.
\label{equ03}
\end{equation}

\noindent where $\rho_0$ is varied from 2 to 8 $\times 10^{8}\ M_{\odot}\cdot\mathrm{pc}^{-3}$ and $r_0 = 10^{\prime\prime}$.

In case of S-stars throughout the whole region which we investigated we can choose only one value of power-law exponent: $\alpha$ = 1.4.

Combination of the above mentioned formulas enabled us to obtain the following expression for the extended mass distribution:

\begin{equation}
M_{ext}(r)=\dfrac{4\pi\rho_0 r_0^\alpha}{3-\alpha}r^{3-\alpha}.
\label{equ04}
\end{equation}

\noindent Note that in \cite{doku15,doku17} the authors used a similar method for estimation of the total dark matter mass near the SMBH at the Galactic Center based on observations of orbital precession of S-stars, and derived an analytical expression for the precession angle in the case of a power-law profile of the dark matter density.

The gravitational potential for extended mass model can be evaluated as \cite{zakh07}:

\begin{equation}
\begin{array} {lcl}
\Phi_{ext}(r) & = &-G\displaystyle\int\limits_r^{r_\infty}\dfrac{M_{ext}(r^\prime)}{r^{\prime 2}}dr^\prime = \\
& & \\
& = & \dfrac{{ - 4\pi {\rho_0} r_0^\alpha G}}{{\left( {3 - \alpha} \right) \left( {2 - \alpha } \right)}}\left( {{r_\infty}^{2 - \alpha} - r^{2 - \alpha}} \right),
\end{array}
\label{equ05}
\end{equation}

\noindent where $r_\infty$ is the outer radius for extended mass distribution of matter. The total gravitational potential is obtained as a sum of Hybrid Palatini potential for SMBH with mass $M_{SMBH}$, and potential for extended matter with mass $M_{ext}(r)$:

\begin{equation}
\Phi_{total}(r)=\Phi(r)+\Phi_{ext}(r).
\label{equ06}
\end{equation}

Modified gravity potential, simillary like GR, give precession arround SMBH. Also, at the center of the Galaxy, around the SMBH, there are invisible sources of mass (clouds of gas, stars and their remnants, and a distributed mass in the form of the diffuse dark matter. This additional invisible sources of mass would cause deviation of the total Newtonian gravitational potential \cite{doku15,doku15b,doku17}. As a result of both effects, the orbits of S-stars would be unclosed and would precess.
If it is assumed that the total potential $\Phi_{total}(r)$ does not differ significantly from Newtonian potential, the perturbed potential has the following form:

\begin{equation}
V_p(r) = \Phi_{total} \left( r \right) - {\Phi_N}\left( r\right)
\begin{array}{*{20}{c}}
;&{{\Phi_N}\left( r \right) = - \dfrac{{GM}}{r}}
\end{array}.
\label{equ07}
\end{equation}

\section{Results and discussion}

In this section we give the estimation of parameters of the Hybrid Palatini gravity model by Schwarzschild precession of S2, S38 and S55 stars, with and without taking into account the bulk mass density distribution of extended matter in the Galactic Center. We assume that the orbital precession of S2, S38 and S55 stars is equal to GR value. The main reason is that the GRAVITY Collaboration detected the orbital precession of the S2 star and showed that it is close to the GR prediction and direction is the same like in GR \cite{abut20}. The second reason is that according to astronomical data fitting in Yukawa gravity model which are presented in the paper \cite{ana21} the orbital precessions of the S38 and S55 stars are also close to the corresponding prediction of GR for these stars.

\subsection{Calculation of orbital precession of S-stars}

A general expression for apocenter shifts for Newtonian potential and small perturbing potential is given as a solution (in Section III Integration of the equations of motion, Chapter 15 Kepler's problem) of problem 3, page 40, equation (1) in the Landau and Lifshitz book \cite{land76}. Assume that a particle moves in slightly perturbed Newtonian potential Adkins and McDonnell \cite{adki07} derived the expression which is equivalent to the above mentioned relation from the Landau and Lifshitz book \cite{land76}, but in an alternative way. It has been showed that the expressions are equivalent and after that they calculated apocenter shifts for several examples of perturbing functions.

According to \cite{adki07}, orbital precession $\Delta\theta$ per orbital period, induced by small perturbations to the Newtonian gravitational potential $\Phi_N(r)=-\dfrac{GM}{r}$ could be evaluated as:

\begin{equation}
\Delta \theta = \dfrac{-2L}{GM e^2}\int\limits_{-1}^1
{\dfrac{z \cdot dz}{\sqrt{1 - z^2}}\dfrac{dV_p\left( z \right)}{dz}},
\label{equ08}
\end{equation}

\noindent while in the textbook \cite{land76} it was given in the form

\begin{equation}
\Delta \theta = \dfrac{2}{GM e}\int\limits_{0}^\pi
\cos \varphi r^2 \dfrac{\partial V_p(r)}{\partial r} d \varphi,
\label{equ09}
\end{equation}

\noindent where $V_p(z)$ is the perturbing potential, $r$ is related to $z$ via: $r = \dfrac{L}{1 + ez}$ in Eq. (\ref{equ08}) (and $r = \dfrac{L}{1 + e \cos \varphi}$ in Eq. (\ref{equ09})), and $L$ being the semilatus rectum of the orbital ellipse with semi-major axis $a$ and eccentricity $e$:

\begin{equation}
L = a\left( {1 - e^2} \right).
\label{equ10}
\end{equation}

Eqs. (\ref{equ08}) and (\ref{equ09}) are equivalent, i.e. Eq. (\ref{equ08}) can be obtained from  Eq. (\ref{equ09}) after substitution: $z=cos \varphi$.

Also, Dokuchaev and Eroshenko \cite{doku15,doku15b,doku17} evaluated relativistic precessions around SMBH in the case of an additional potential due to a presence of dark matter. The precession angle per orbital period is expressed analytically using hypergeometric function \cite{doku15,doku15b,doku17}:

\begin{equation}
\delta \theta = - \dfrac{4 \pi^2 \rho_0 r_0^\alpha L^{3-\alpha}}{(1 - e)^{4-\alpha} M_{SMBH}} \, _2F_1 \left( 4 - \alpha, \dfrac{3}{2}; 3; -\dfrac{2e}{1 - e} \right),
\label{equ11}
\end{equation}

\noindent where $_2F_1$ is the hypergeometric function. This expression is in good agreement with the corresponding expression given in the Landau and Lifshitz book \cite{land76}. More details are given in references \cite{doku15,doku15b,doku17}.
Besides, if one takes the expressions for precession from the books by Danby \cite{danb62} (Chapter 11 equation 11.5.13) and by Murray and Dermott \cite{murr00} (Chapter 2, equation 2.165.) one can obtain the same equations as the above Eq. (\ref{equ08}).

\begin{figure}[ht!]
\centering
\includegraphics[width=0.60\textwidth]{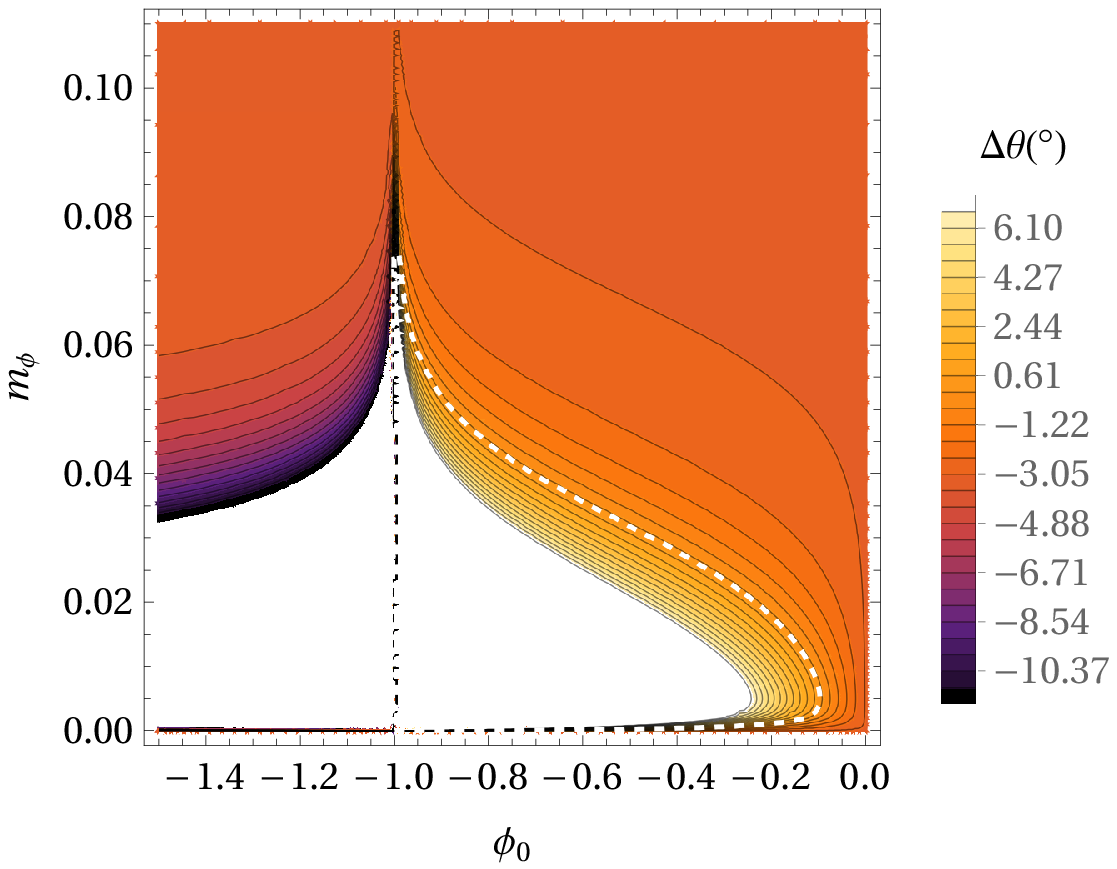} \\
\vspace{0.5cm}
\includegraphics[width=0.60\textwidth]{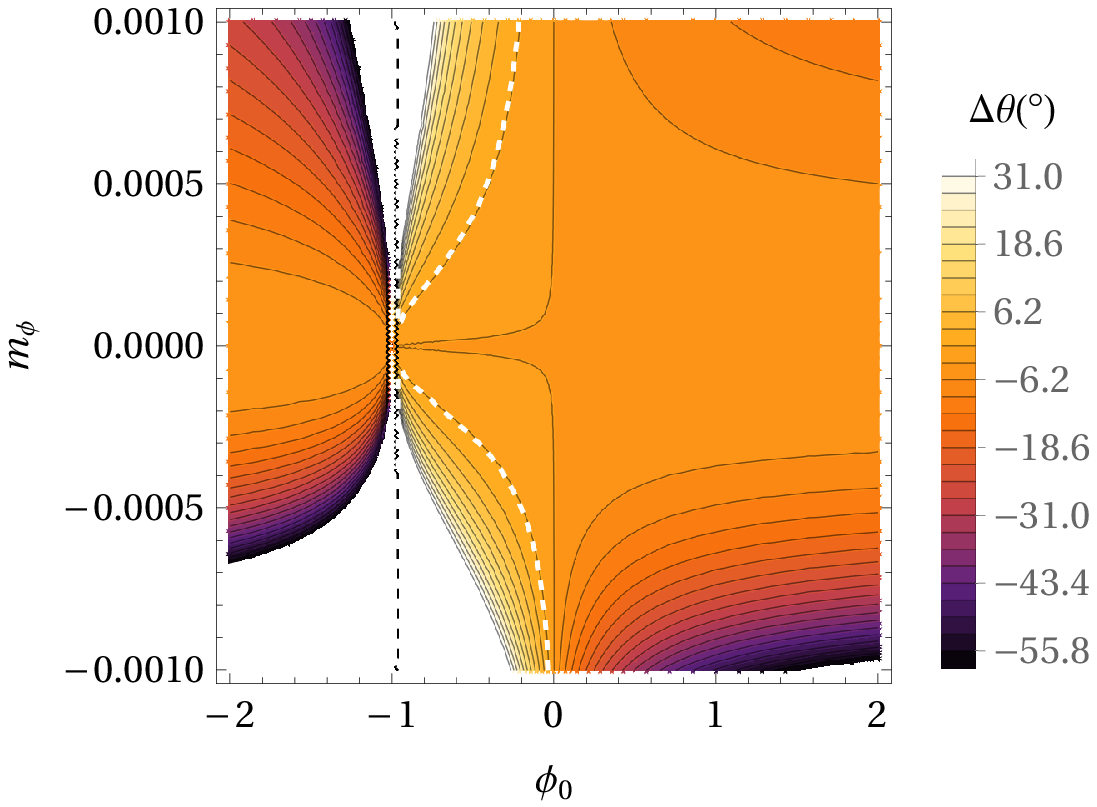}
\caption{The precession per orbital period for $\phi_0$ - $m_\phi$ parameter space in the case of Hybrid Palatini gravity potential with extended mass distribution in case of S2 star. The mass density distribution of extended matter is $\rho_0$ = $2 \times 10^8 M_\odot \mathrm{pc^{-3}}$. With a decreasing value of precession angle colors are darker. Parameter $m_\phi$ is expressed in AU$^{-1}$. White dashed line depicts the locations in parameter space where precession angle has the same value as in GR ($0^\circ.18$). Lower panel represents the same like the upper panel but for smaller values of $m_\phi$ parameter.}
\label{fig01}
\end{figure}

\begin{figure}[ht!]
\centering
\includegraphics[width=0.60\textwidth]{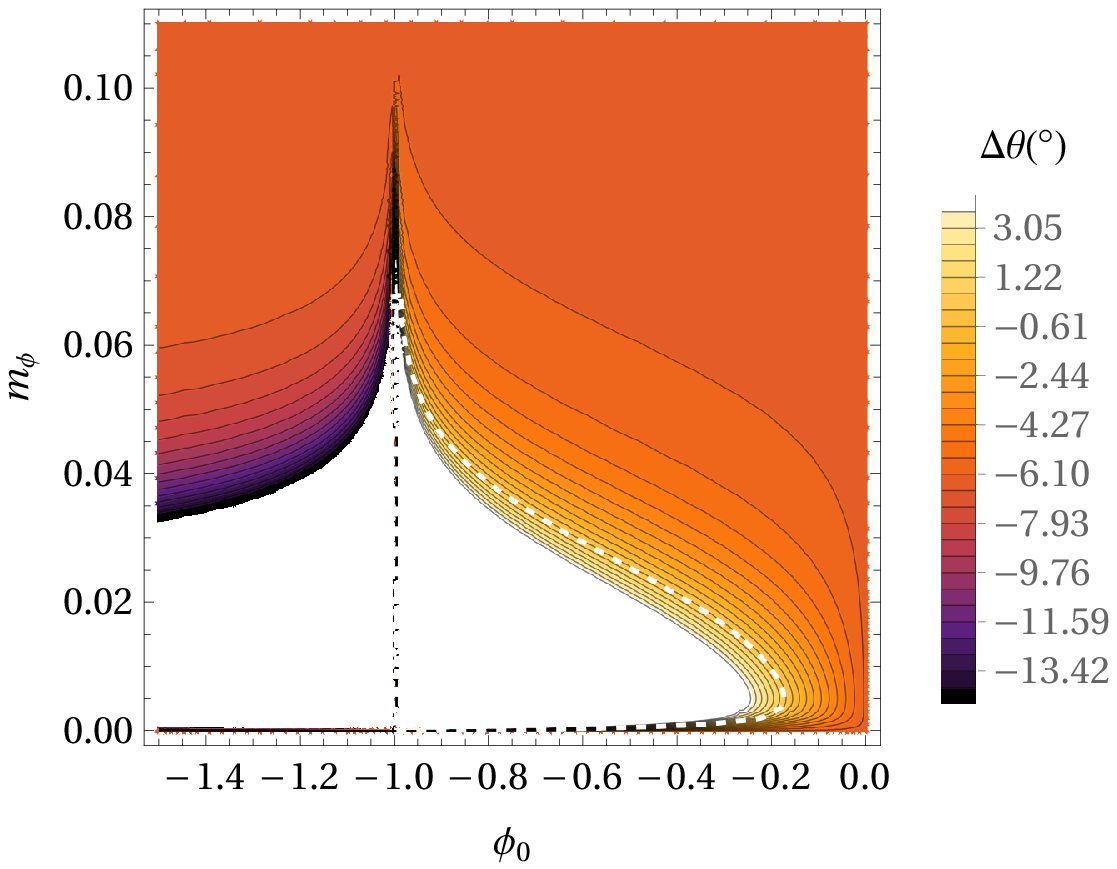} \\
\vspace{0.5cm}
\includegraphics[width=0.60\textwidth]{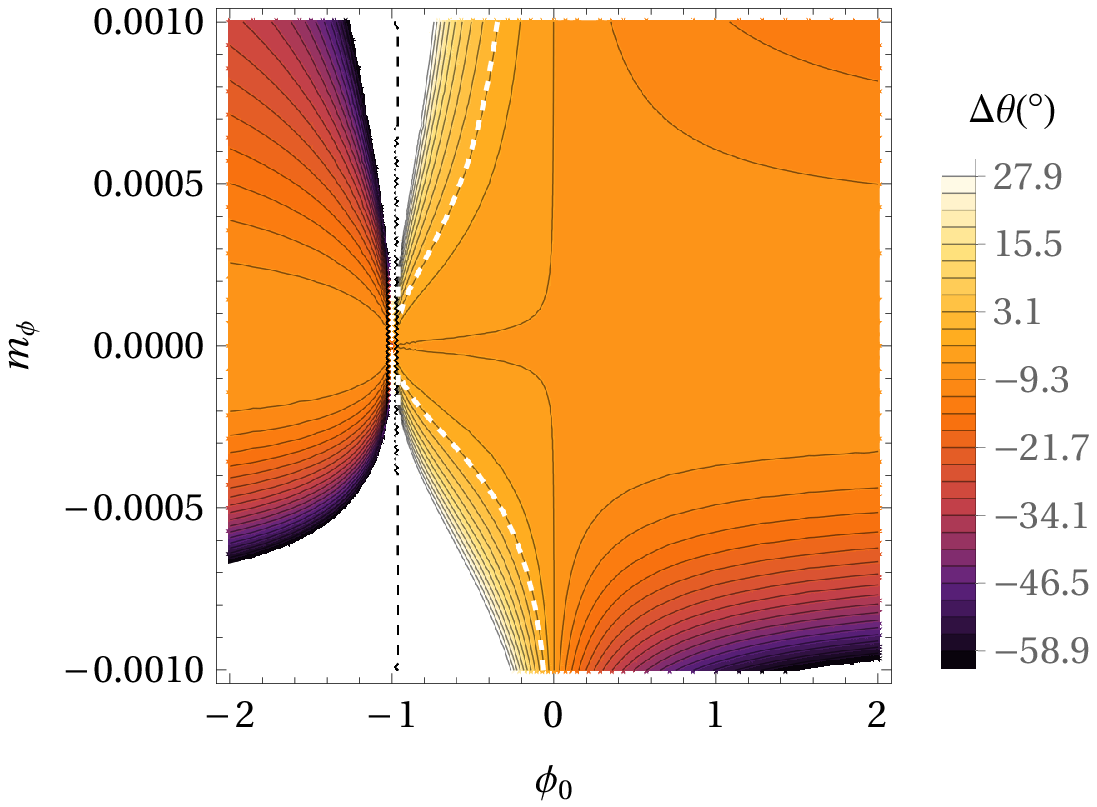}
\caption{The same as in Fig. \ref{fig01} but for the values of the mass density distribution of extended matter $\rho_0$ = $4 \times 10^8 M_\odot \mathrm{pc^{-3}}$. Lower panel represents the same like the upper panel but for smaller values of $m_\phi$ parameter.}
\label{fig02}
\end{figure}

\begin{figure}[ht!]
\centering
\includegraphics[width=0.60\textwidth]{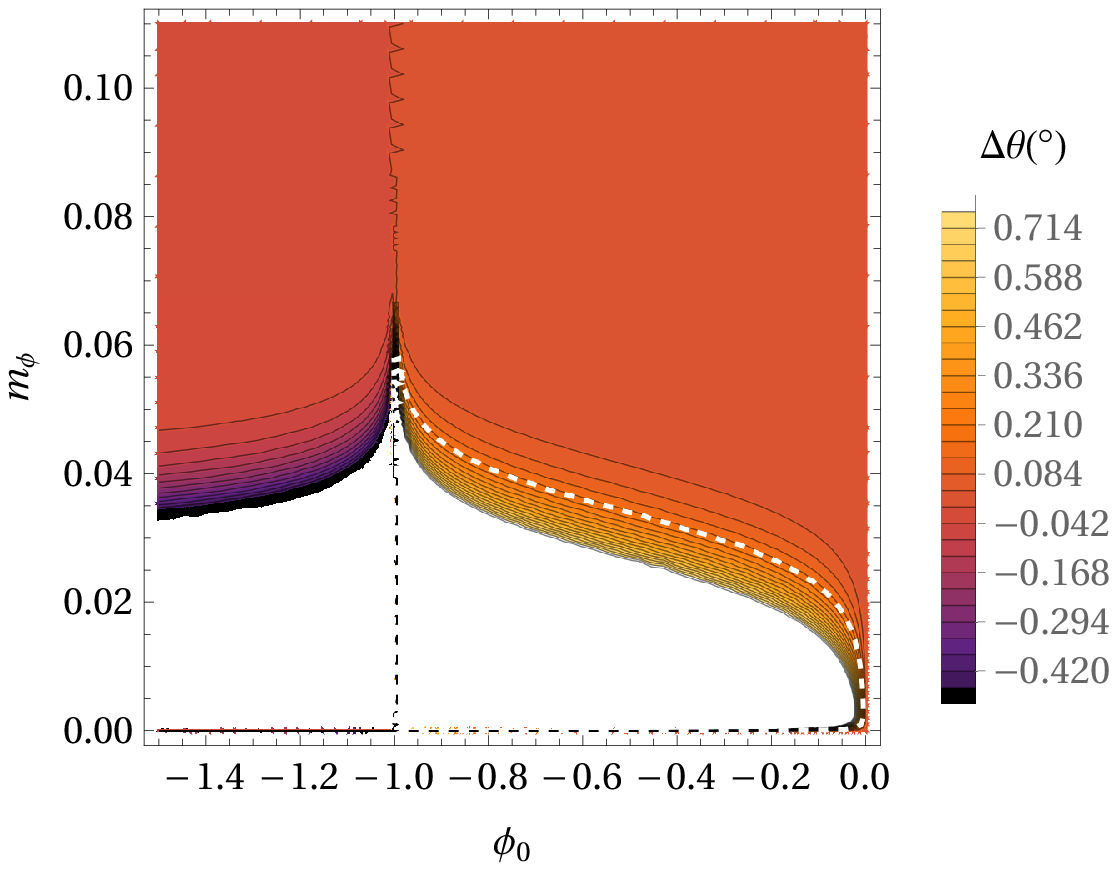} \\
\vspace{0.5cm}
\includegraphics[width=0.60\textwidth]{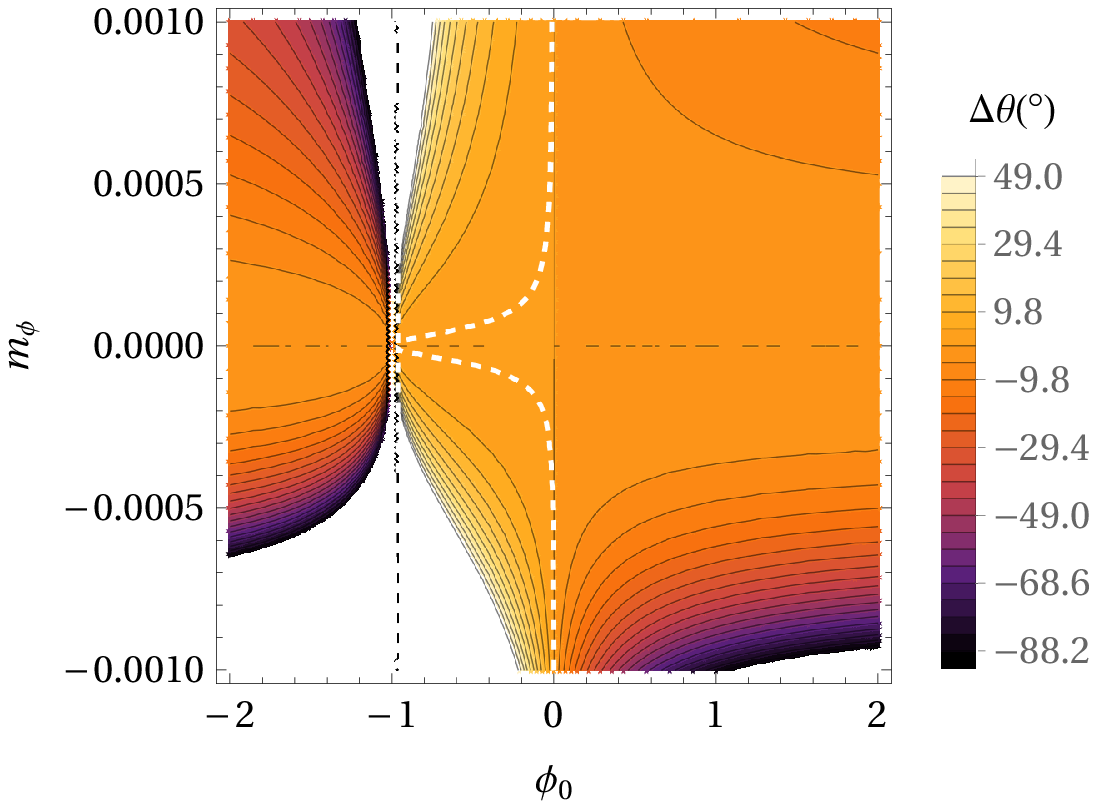}
\caption{The precession per orbital period for $\phi_0$ - $m_\phi$ parameter space in the case of Hybrid Palatini gravity potential without extended mass distribution in case of S38 star. With a decreasing value of precession angle colors are darker. Parameter $m_\phi$ is expressed in AU$^{-1}$. White dashed line depicts the locations in parameter space where precession angle has the same value as in GR ($0^\circ.11$). Lower panel represents the same like the upper panel but for smaller values of $m_\phi$ parameter.}
\label{fig03}
\end{figure}

\begin{figure}[ht!]
\centering
\includegraphics[width=0.60\textwidth]{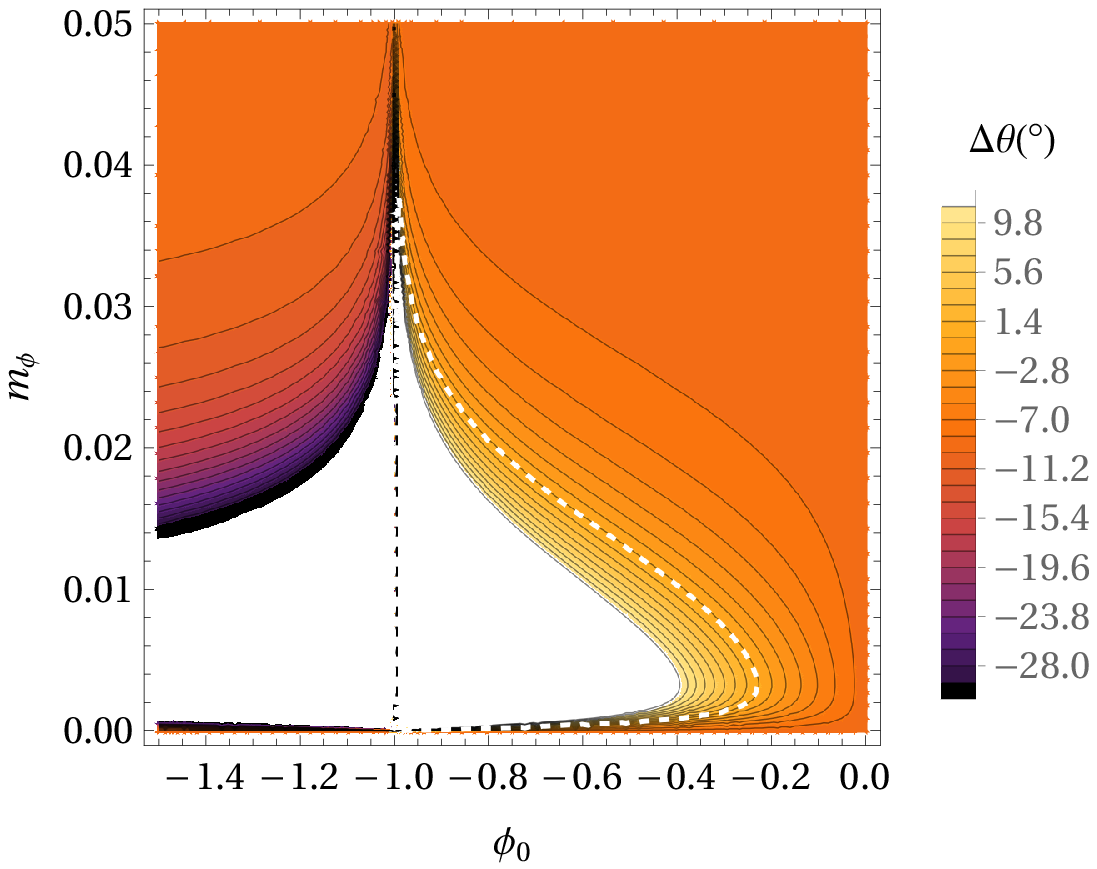} \\
\vspace{0.5cm}
\includegraphics[width=0.60\textwidth]{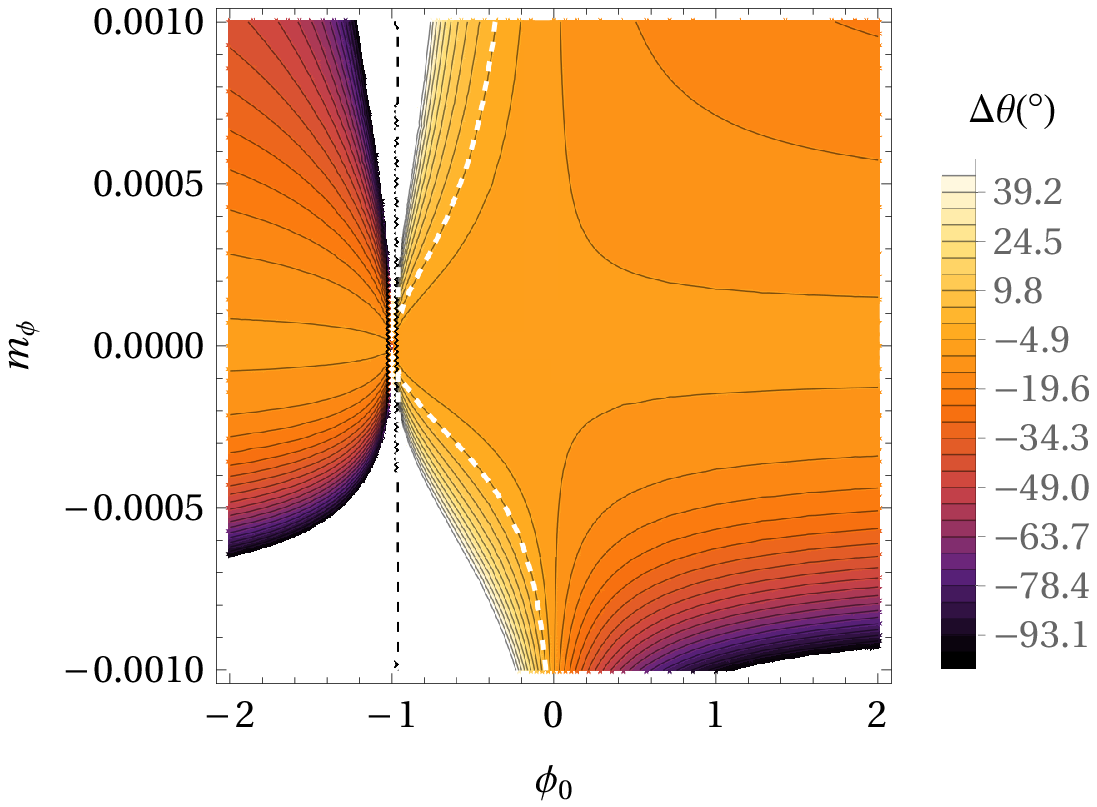}
\caption{The same as in Fig. \ref{fig03}, but for the mass density distribution $\rho_0$ = $4 \times 10^8 M_\odot \mathrm{pc^{-3}}$.  Lower panel represents the same like the upper panel but for smaller values of $m_\phi$ parameter.}
\label{fig04}
\end{figure}

\begin{figure}[ht!]
\centering
\includegraphics[width=0.60\textwidth]{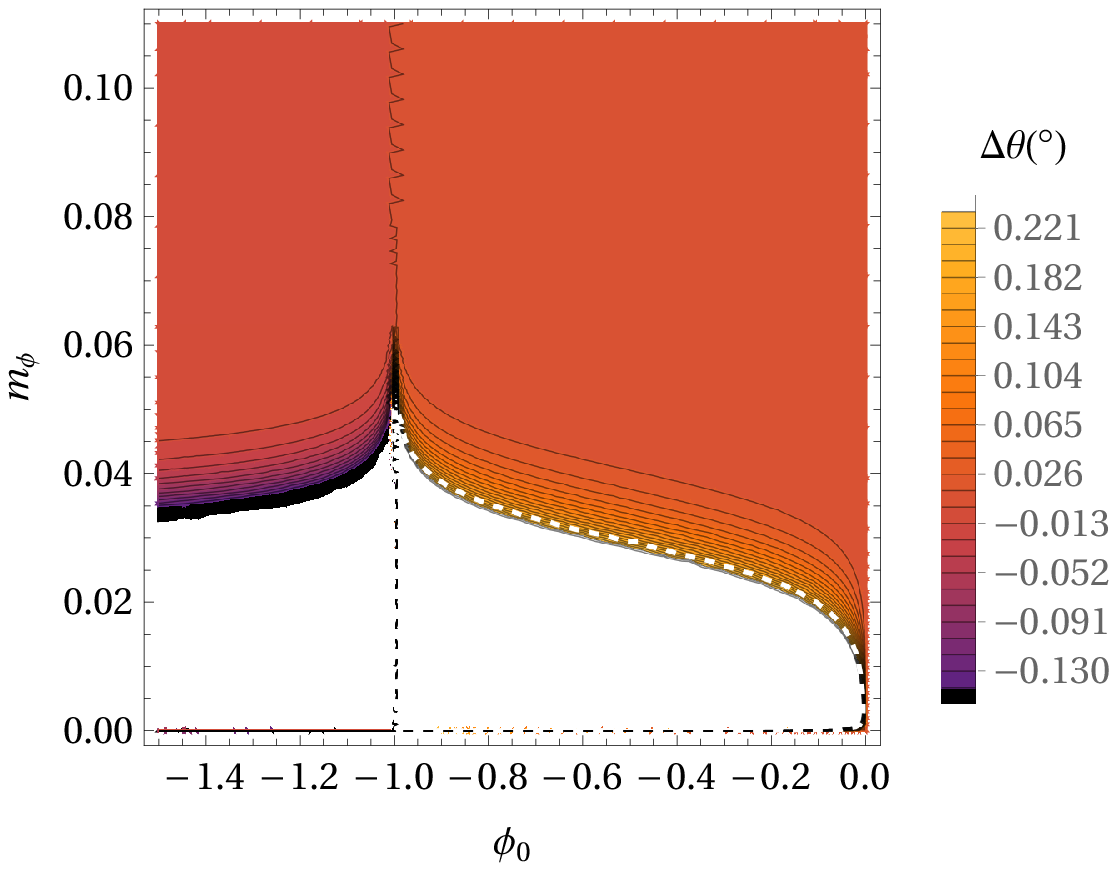} \\
\vspace{0.5cm}
\includegraphics[width=0.60\textwidth]{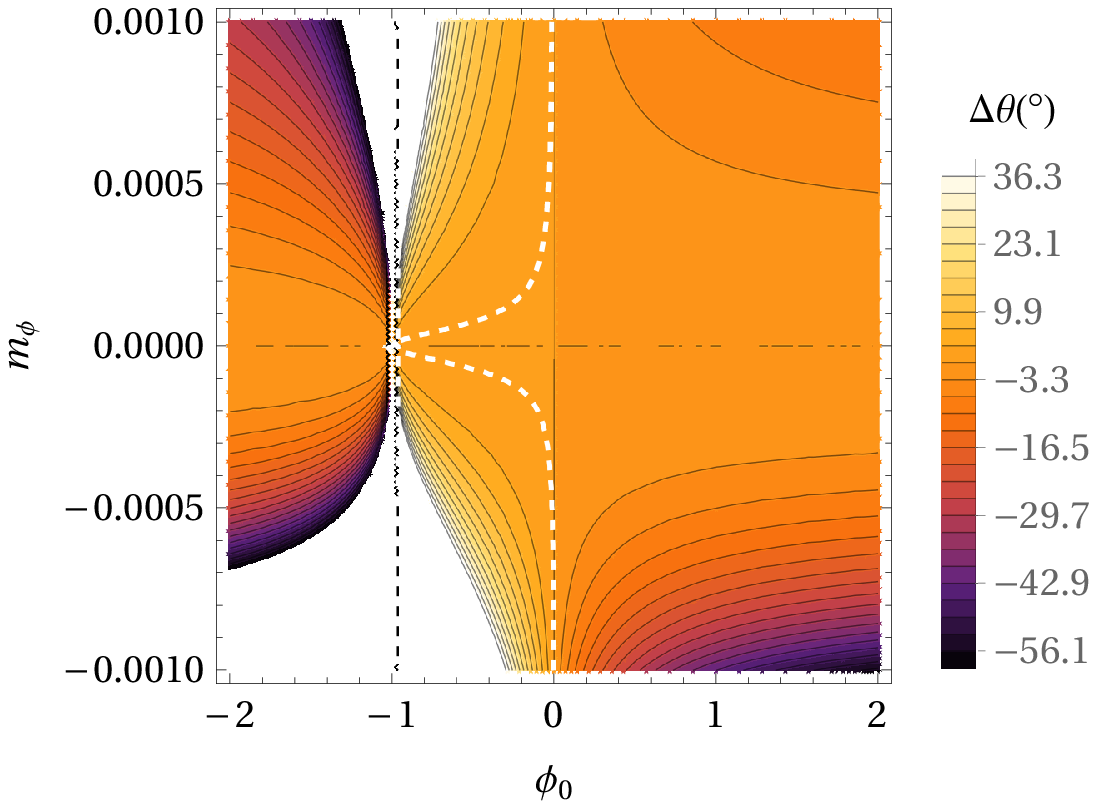}
\caption{The precession per orbital period for $\phi_0$ - $m_\phi$ parameter space in the case of Hybrid Palatini gravity potential without extended mass distribution in case of S55 star. With a decreasing value of precession angle colors are darker. Parameter $m_\phi$ is expressed in AU$^{-1}$. White dashed line depicts the locations in parameter space where precession angle has the same value as in GR ($0^\circ.10$). Lower panel represents the same like the upper panel but for smaller values of $m_\phi$ parameter.}
\label{fig05}
\end{figure}

\begin{figure}[ht!]
\centering
\includegraphics[width=0.60\textwidth]{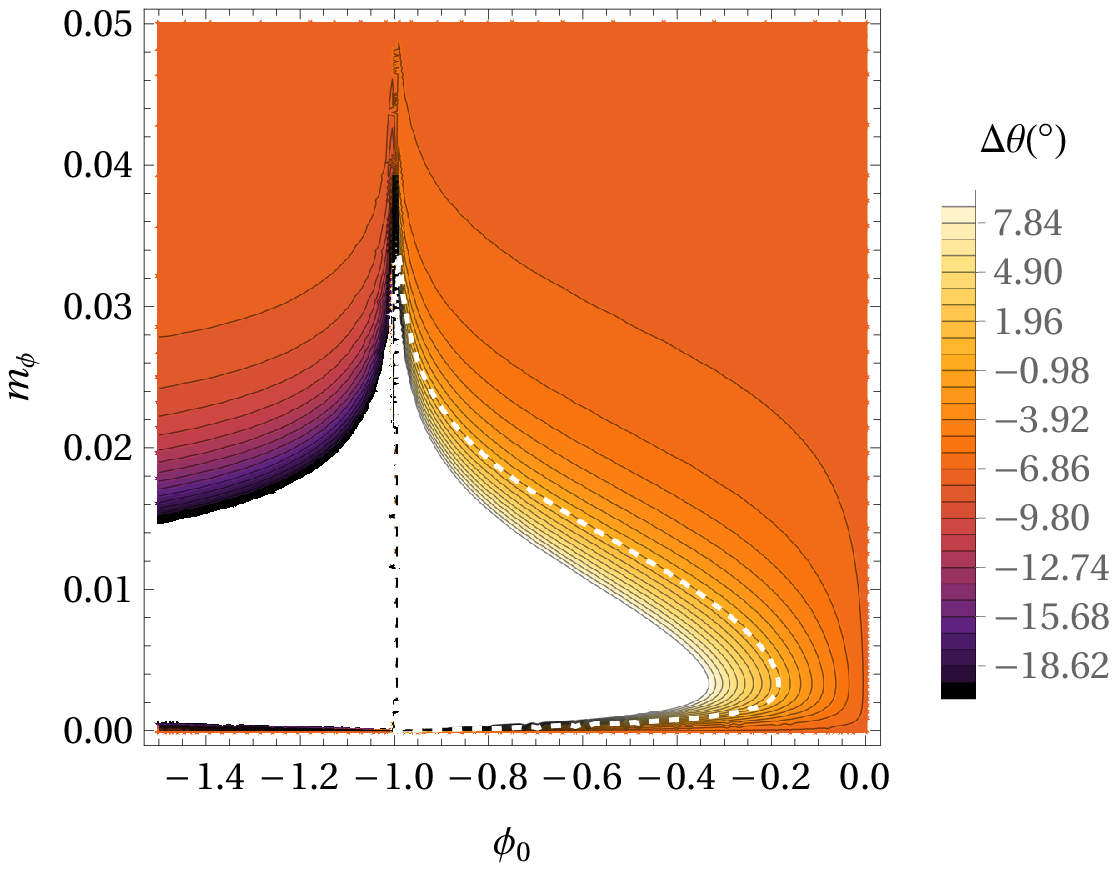} \\
\vspace{0.5cm}
\includegraphics[width=0.60\textwidth]{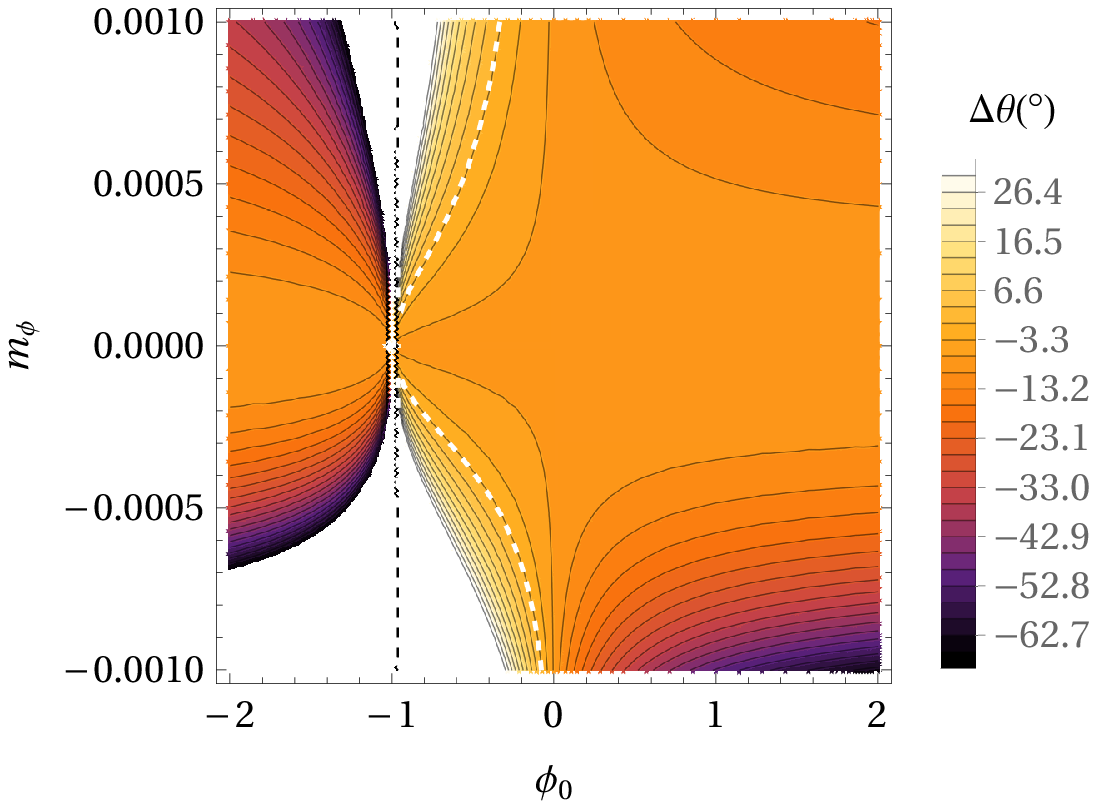}
\caption{The same as in Fig. \ref{fig05}, but for the mass density distribution $\rho_0$ = $4 \times 10^8 M_\odot \mathrm{pc^{-3}}$.  Lower panel represents the same like the upper panel but for smaller values of $m_\phi$ parameter.}
\label{fig06}
\end{figure}

To calculate the precession of the S2, S38 and S55 stars in Hybrid Palatini modified gravity, we assumed that the perturbed potential is of the form:

\begin{equation}
V_p(r) = \Phi(r)+\Phi_{ext}(r) - {\Phi_N}(r); \quad {{\Phi_N}(r) = - \dfrac{{GM}}{r}},
\label{equ12}
\end{equation}

\noindent and it can be used to calculate the precession angle according to the Eq. (\ref{equ08}):

In order to investigate the parameters of Hybrid Palatini gravity, which in the case of the extended mass distribution give the same orbital precession as GR, we graphically presented Eq. (\ref{equ08}) by adopting different values of the extended mass density $\rho_0$, and for three different S-stars. In that way we created the below Figs. \ref{fig01}-\ref{fig06} showing the dependence of orbital precession angle $\Delta \theta$ on the gravity parameters $\phi_0$ and $m_\phi$ for several extended mass densities $\rho_0$ and for the following three S-stars: S2, S38 and S55.
The observed quantities which are used in this paper are the parameters of the central SMBH in our Galaxy, as well as the orbital elements for the mentioned stars.

For our calculation we used the results presented in \cite{gill17}, according to which mass of the SMBH of the Milky Way is $M_{SMBH} = 4.3\times 10^6\ M_\odot$, semi-major axis of the S2 star orbit is $a=0.''1255$, and its eccentricity is $e = 0.8839$; semi-major axis of the S38 star orbit is $a=0.''1416$, and its eccentricity is $e = 0.8201$; and semi-major axis of the S55 star orbit is $a=0.''1078$, and its eccentricity is $e = 0.7209$.

Fig. \ref{fig01} shows the precession per orbital period for $\phi_0$ - $m_\phi$ parameter space in the case of Hybrid Palatini gravity potential with extended mass distribution in case of S2 star. The mass density distribution of extended matter is $\rho_0$ = $2 \times 10^8 M_\odot \mathrm{pc^{-3}}$. White dashed line depicts the locations in parameter space where precession angle has the same value as in GR for S2 star ($0^\circ.18$). It can be shown that precession of orbit in Hybrid Palatini potential is in the same direction as in GR \cite{bork16}, but extended mass distribution produce a contribution to precession in opposite direction \cite{jova21}. According to Fig. \ref{fig01} and formulas for potential in Modified Hybrid Palatini gravity (see denominator in eq. \ref{equ01}), parameter $\phi_0$ is between -1 (vertical asymptote) and 0. If $\phi_0=0$ the Hybrid Palatini potential reduces to the Newtonian one. Maximal value for $m_\phi$ is about 0.075 AU$^{-1}$ and for $m_\phi$ near 0.005 AU$^{-1}$ maximal value for $\phi_0$ is obtained and it is around -0.1 (see upper panel). We can see from lower panel that $m_\phi$ also can take negative values, but when $m_\phi$ become less than -0.0001 AU$^{-1}$ parameter $\phi_0$ become very near 0 and the Hybrid Palatini potential reduces to the Newtonian one.

Fig. \ref{fig02} represents the same as Fig. \ref{fig01} but for the values of the mass density distribution of extended matter $\rho_0$ = $4 \times 10^8 M_\odot \mathrm{pc^{-3}}$. We can notice the similar tendency like in previous cases regarding dependence of shape of dashed curve with respect to the values of parameters $m_\phi$ and $\phi_0$. Maximal value for $m_\phi$ is about 0.065 AU$^{-1}$ and for $m_\phi$ near 0.005 AU$^{-1}$ maximal value for $\phi_0$ is obtained and it is around -0.17. If we compared Figs. \ref{fig01} and \ref{fig02} with the corresponding Fig. 4 from paper \cite{bork21b} where we did not take into account extended mass distribution (maximal value for $m_\phi$ is about 0.10 AU$^{-1}$ and for $m_\phi$ near 0.005 AU$^{-1}$ maximal value for $\phi_0$ is obtained and it is around -0.01) we can conclude that the mass density distribution of extended matter $\rho_0$ has strong influence on the gravity parametar $m_\phi$ and value of precession angle per orbital period for S2 star. If we increase value of $\rho_0$, we obtain the decrease of corresponding values of parameters $m_\phi$ and $\phi_0$.

Fig. \ref{fig03} shows the precession per orbital period for $\phi_0$ - $m_\phi$ parameter space in the case of Hybrid Palatini gravity potential without extended mass distribution in case of S38 star. White dashed line depicts the locations in parameter space where precession angle has the same value as in GR for S38 star ($0^\circ.11$). Maximal value for $m_\phi$ is about 0.06 AU$^{-1}$ and for $m_\phi$ near 0.005 AU$^{-1}$ maximal value for $\phi_0$ is obtained and it is around -0.01. According to the lower panel, we can see that $m_\phi$ also can take negative values. Fig. \ref{fig04} represent the same as Fig. \ref{fig03}, but for the mass density distribution $\rho_0$ = $4 \times 10^8 M_\odot \mathrm{pc^{-3}}$. Maximal value for $m_\phi$ is less than 0.04 AU$^{-1}$ and for $m_\phi$ near 0.005 AU$^{-1}$ maximal value for $\phi_0$ is obtained and it is less than -0.2.

Figs. \ref{fig05} and \ref{fig06} represent the same as Figs. \ref{fig03} and \ref{fig04}, but for the S55 star (precession angle in GR is $0^\circ.10$). If we compare the estimated parameters of the Hybrid Palatini gravity model of S2 star with S38 and S55 stars for the same value of $\rho_0$, it can be seen that results are slightly different, i.e. the obtained values for parameters $\phi_0$ and $m_\phi$ are not the same, but they are very close. It seems that parameters of the Hybrid Palatini gravity depend on the scale (values of semi-major axes).

According to Figs. \ref{fig01} - \ref{fig06} the mass density distribution of extended matter has significant influence on the values of precession angle and of parameters $\phi_0$ and $m_\phi$. Also, we can notice that it is not possible to evaluate $\phi_0$ and $m_\phi$ in a unique way, if we consider only following two conditions: 1) the orbital precession is prograde like in GR, and 2) the value of precession angle is like in GR. In fact, we obtained lines in $\phi_0$ - $m_\phi$ parameter space and points of these lines have coordinates $\phi_0$ and $m_\phi$ which fulfilled above mentioned two requests. If we want to obtain only one unique value of parameters $\phi_0$ and $m_\phi$ we need additional independent set of observations to combine with these obtained sets of points ($\phi_0$, $m_\phi$).

This paper is continuation of our previous research \cite{bork21}, but we extended our research on the following points:
\begin{enumerate}[label=(\roman*),nosep]
\item In this study, we estimated parameters of the Hybrid Palatini gravity model with the Schwarzschild precession of S-stars. Beside S2 star, here we for the first time take into account S38 and S55 stars also. If we compare the estimated parameters of the Hybrid Palatini gravity model of S2 star with S38 and S55 stars, it can be seen that parameters of the Hybrid Palatini gravity depend on the scale of a gravitational system, which in this case is the semi-major axis of a stellar orbit.
\item Also, in this paper we considered orbital precession of the mentioned stars due to additional contribution to the gravitational potential from a bulk distribution of matter. We take into account the different values of bulk mass density distribution of extended matter in the Galactic Center and analyze their influence on values of parameters $m_\phi$ and $\phi_0$ of Hybrid Palatini gravity model. We conclude that the mass density distribution of extended matter has significant influence on the values of precession angle and of modified gravity parameters. For higher values of $\rho_0$ we obtained lower values of gravity parameters $m_\phi$ and $\phi_0$. This paper is also extension of our previous paper where we investigate gravity parameters of Yukawa theory and how they change under different values of bulk mass density distribution of extended matter \cite{jova21}. In this paper we apply the same procedure but for parameters of Hybrid Palatini gravity model and we extended it to S38 and S55 stars.
\item We believe that besides the most often used S2 star, S38 and S55 stars are also excellent candidates for probing the gravitational potential around central SMBH and could be aslo very useful for evaluating accurate parameters of different alternative gravity models.
\item In our previous paper \cite{bork16}, where we constrained parameters of Hybrid Palatini gravity, we used observational data from VLT and Keck collaborations. The results were obtained by fitting the simulated orbits of S2 star to its observed astrometric positions. Observational data were obtained with relatively large errors, especially at the first stage of monitoring (data were collected for decades). In this paper we did not fit the observational data, but instead we only assumed that the orbital precession of S2 star is equal to the corresponding value predicted by GR because recently the GRAVITY Collaboration claimed that they detected the orbital precession of the S2 star and showed that it is close to the GR prediction \cite{abut20}. Also, we extended our analysis to the stars S38 and S55 stars because astronomical data analysis of their orbits showed that also in these cases orbital precession is close to the GR prediction \cite{ana21}.
\end{enumerate}

\section{Conclusions}

In this study, we estimated parameters of the Hybrid Palatini gravity model with the Schwarzschild precession of S2, S38 and S55 stars. We estimated parameters with and without taking into account case of bulk mass distribution near Galactic Center. In this study we were not fitting observation data, but instead we assume that the Schwarzschild orbital precessions of S2, S38 and S55 stars are the same like in of GR, i.e. $0^\circ.18$, $0^\circ.11$ and $0^\circ.10$ per orbital period, respectively. We introduce this approximation, since the observed precession angle of S2 star is very close to GR prediction \cite{abut20} and according the paper \cite{ana21} where author analyzed observation data in the framework of Yukawa gravity and conclude that the orbital precessions of the S38 and S55 stars are in good agreement to the corresponding prediction of GR for these stars. Also, we have second request, i. e. that we should recover prograde orbital precession of S-stars, like in GR. Our findings indicate that:

\begin{enumerate}
\item Modified Hybrid Palatini gravity parameter $\phi_0$ is between -1 (vertical asymptote) and 0. If $\phi_0=0$ the Hybrid Palatini gravity potential reduces to the Newtonian one.	
\item For the Hybrid Palatini gravity model (described with two parameters) it is not possible to evaluate both parameters in a unique way, if we consider only the conditions that orbital precession is prograde like in GR, and that the value of precession angle is like in GR. Instead of that we obtained lines in $\phi_0$ - $m_\phi$ parameter space. Points of these lines have coordinates $\phi_0$ and $m_\phi$ which fulfilled our two requests (value of precession like in GR and precession is prograde like in GR). White dashed line depicts the locations in parameter space of these points. If we want to obtain only one value of parameters $\phi_0$ and $m_\phi$, we need to combine obtained sets of ($\phi_0$, $m_\phi$) with some additional independent set of observations. 
\item The mass density distribution of extended matter has significant influence on the values of precession angle and of modified gravity parameters. Higher values of $\rho_0$ decrease the corresponding values of parameters $m_\phi$ and $\phi_0$. 
\item Our analysis shows that precession of orbit in Hybrid Palatini potential is in the same direction as in GR, but extended mass distribution produces a contribution to precession in opposite direction. It means that, for higher mass densities, in order to obtain the same orbital precession as in GR, one has to take the significantly different values of Hybrid Palatini gravity parameters. In the case when $\phi_0=0$ Hybrid Palatini gravitational potential reduces to the Newtonian one. However, in order to compensate the effects of extended mass distribution on orbital precession and to obtain the same precession as in GR, $\phi_0$ has to be larger by absolute value, causing the larger deviation of Hybrid Palatini gravitational potential with respect to the Newtonian one.
\item If we compare the estimated parameters of the Hybrid Palatini gravity model of S2 star with S38 and S55 stars, it can be seen that results are slightly different, i.e. the obtained values for parameters of gravity models are not the same, but they are very close. It seems that the parameters of the Hybrid Palatini gravity depend on the scale of a gravitational system, which in this case is the semi-major axis of a stellar orbit, in contrast to GR which is the scale-invariant theory of gravitation. Therefore, we believe that this behaviour originate from the deviation of modified gravity from general relativity.
\end{enumerate}

It is very important to investigate gravity in the vicinity of very massive compact objects like Sgr A$^\ast$, because the environment around these objects is drastically different from that in the Solar System framework or at extragalactic and cosmological scales. Also, the precession of the S stars is a unique opportunity to test gravity at the sub-parsec scale of a few thousand AU because these stars are bright stars and periods of these stars are relatively short. We believe that it is useful to evaluate parameters of different alternative modified gravity theories in the vicinity of SMBH with and without extended mass distribution in metric and Palatini approach. There are various approaches to construction of the modified gravity theories. In general, one can classify most efforts as modified gravity or introducing some exotic matter like dark matter and dark energy. The truth, as, usual, may lie in between \cite{alle05}.

We hope that using this method and more precise astronomical data will help to evaluate accurate parameters of different alternative gravity models and to obtain gravitational potential at the Galactic Center.

\end{paracol}
\begin{paracol}{2}
\switchcolumn

\authorcontributions{All coauthors participated in writing, calculation and discussion of obtained results.}

\acknowledgments{This work is supported by Ministry of Education, Science and Technological Development of the Republic of Serbia. PJ wishes to acknowledge the support by this Ministry through the project contract No. 451-03-9/2021-14/200002. The authors also wish to thank the Center for mathematical modeling and computer simulations in physics and astrophysics of Vin\v{c}a Institute of Nuclear Sciences.}

\conflictsofinterest{The authors declare no conflict of interest.} 

\abbreviations{Abbreviations}{The following abbreviations are used in this manuscript:\\
	
\noindent 
\begin{tabular}{@{}ll}
GR & General Relativity \\
LT & Lense-Thirring \\
SMBH & Supermassive black hole \\
\end{tabular}}

\appendixtitles{yes} 
\appendixstart
\appendix
\section{Hybrid Palatini gravity model}

It is important to note that theoretical studies in this field commonly assume $c$ = $G$ = 1 units. However, for practical purposes, i.e. for comparisons with the astronomical observations, it is necessary to recast gravitation potential in appropriate units. Thus, here we derive gravitation potential in weak field limit of Hybrid Palatini gravity in the form convenient for this purpose.

The action, proposed in the papers by Capozziello et al. (2012,2013) \cite{capo12,capo13}, Harko et al. (2012) \cite{hark12}, Borka et al. (2016) \cite{bork16}, is given by:

\begin{equation}
S =\dfrac{1}{2\kappa}\int{d^4 x\sqrt{-g} \left[R + \phi \mathcal{R} - V(\phi) + 2 \kappa L_m \right]},
\label{equA01}
\end{equation}

\noindent where $\kappa = \dfrac{8\pi G}{c^4}$, $R$ is Ricci scalar, $\mathcal{R} = g^{\mu\nu} \mathcal{R}_{\mu \nu}$ presents Palatini curvature with the independent connection $\widetilde{\Gamma}^\lambda_{\mu \nu}$, $L_m$ is density Lagrangian, $g$ is determinant of $g_{\mu\nu}$.

Palatini curvature is given by the following equations, with the scalar field $\phi$ and potential $V(\phi)$:

\begin{equation}
\mathcal{R}_{\mu\nu} \equiv \widetilde{\Gamma}_{\mu\nu,\alpha}^{\alpha} - \widetilde{\Gamma}_{\mu\alpha,\nu}^{\alpha} + \widetilde{\Gamma}_{\alpha\lambda}^{\alpha}\widetilde{\Gamma}_{\mu\nu}^{\lambda} - \widetilde{\Gamma}_{\mu\lambda}^{\alpha}\widetilde{\Gamma}_{\alpha\nu}^{\lambda}
\label{equA02}
\end{equation}

\begin{equation}
\widetilde{\Gamma}_{\mu\nu}^{\lambda} = \dfrac{1}{2}\widetilde{g}^{\lambda\sigma}(\widetilde{g}_{\mu\sigma,\nu} + \widetilde{g}_{\nu\sigma,\mu} - \widetilde{g}_{\mu\nu,\sigma})
\label{equA03}
\end{equation}

\begin{equation}
\widetilde{g}_{\lambda\sigma} = g_{\lambda\sigma}F(\mathcal{R}).
\label{equA04}
\end{equation}

\noindent Combination of the Eqs. (\ref{equA02}), (\ref{equA03}) and (\ref{equA04}) resulted in the equation:

\begin{eqnarray}
\widetilde{\Gamma}_{\mu\nu}^{\lambda} &=& \dfrac{g^{\lambda\sigma}}{2F(\mathcal{R})} \Big(g_{\mu\sigma,\nu}F(\mathcal{R}) + g_{\nu\sigma,\mu}F(\mathcal{R}) - g_{\mu\nu,\sigma}F(\mathcal{R}) + \nonumber \\ 
&& + g_{\mu\sigma}F(\mathcal{R})_{,\nu} + g_{\nu\sigma}F(\mathcal{R})_{,\mu} - g_{\mu\nu}F(\mathcal{R})_{,\sigma} \Big).  
\label{equA05}
\end{eqnarray}

Substitution of Eq. (\ref{equA05}) into Eq. (\ref{equA02}) enabled to obtain the expression for Palatini curvature:

\begin{equation}
\mathcal{R}_{\mu\nu} = R_{\mu\nu} + \dfrac{3 \nabla_\mu F(\mathcal{R}) \nabla_\nu F(\mathcal{R})}{2F(\mathcal{R})^2} - \dfrac{\nabla_\mu \nabla_\nu F(\mathcal{R})}{F(\mathcal{R})} - \dfrac{g_{\mu\nu}}{2} \dfrac{\Box F(\mathcal{R})}{F(\mathcal{R})}.
\label{equA06}
\end{equation}

The action is varied respectively to the metric $g_{\mu\nu}$, scalar field $\phi$ and connection $\widetilde{\Gamma}_{\mu \nu}^{\lambda}$, which leads to following equations:

\begin{equation}
R_{\mu \nu} + \phi \mathcal{R}_{\mu \nu} - \dfrac{1}{2} g_{\mu \nu} [R + \phi \mathcal{R} - V({\phi})] = \kappa T_{\mu \nu}
\label{equA07}
\end{equation}

\begin{equation}
\mathcal{R} - V'(\phi) = 0
\label{equA08}
\end{equation}

\begin{equation}
\widetilde{\nabla}_{\alpha}(\sqrt{-g}\phi g^{\mu\nu}) = 0.
\label{equA09}
\end{equation}

Palatini connection is represented by Eq. (\ref{equA09}) \cite{wu18}, which is obtained by varied action with respect to the relation $\widetilde{\Gamma}_{\mu\nu}^{\lambda}$, by keeping the metric constant $g^{\mu\nu}$. Eq. (\ref{equA09}) implied that the function $F(R) = \phi$, so the Palatini Tensor and Palatini scalar are given by following equations: 

\begin{equation}
\mathcal{R}_{\mu \nu} = R_{\mu \nu} + \dfrac{3\partial_\mu\phi \partial_\nu\phi}{2\phi^2} - \dfrac{\nabla_\mu \nabla_\nu \phi}{\phi} - \dfrac{g_{\mu\nu}}{2} \dfrac{\Box \phi}{\phi},
\label{equA10}
\end{equation}

\begin{equation}
\mathcal{R} = R + \dfrac{3\partial_\mu\phi \partial^\mu\phi}{2\phi^2} - \dfrac{3 \ \Box\phi}{\phi}.
\label{equA11}
\end{equation}

The trace of Eq. (\ref{equA07}) is given in the next relation:

\begin{equation}
R + \kappa T = 2V(\phi) - \phi V_{\phi}, \quad V'(\phi) = V_{\phi}.
\label{equA12}
\end{equation}

Combination of Eqs. (\ref{equA08}), (\ref{equA10}), (\ref{equA12}), and Eq. (\ref{equA07}), enabled obtaining of the metric field equations:

\begin{equation}
(1+\phi) R_{\mu\nu} = \kappa (T_{\mu\nu} - \dfrac{1}{2}g_{\mu\nu}T) + \dfrac{1}{2} g_{\mu\nu}(V(\phi) + \Box \phi) + \nabla_\mu \nabla_\nu \phi - \dfrac{3\partial_\mu\phi \partial_\nu\phi}{2\phi}. 
\label{equA13}
\end{equation}

\noindent and the trace of Eq. (\ref{equA13}) is:

\begin{equation}
(1+\phi)R = - \kappa T + 2(V(\phi) + \Box \phi) + \Box\phi - \dfrac{3\partial_\mu\phi \partial^\mu\phi}{2\phi}.
\label{equA14}
\end{equation}

The scalar field equation is obtained by combination of the Eqs. (\ref{equA12}) and (\ref{equA14}):

\begin{equation}
- \Box \phi + \dfrac{\partial_\mu\phi \partial^\mu\phi}{2\phi} + \dfrac{\phi}{3} (2V(\phi) - (1 + \phi)V_{\phi}) = \dfrac{\phi \kappa T}{3}.
\label{equA15}
\end{equation}

We can see that scalar field is governed by the second order evolution equation, which is an effective Klein-Gordon equation. 

\subsection{Equations for Newtonian limit}

In order to derive the Newtonian limit, it is common to write metric $g_{\mu\nu}$ as a sum of Minkowski metric $\eta_{\mu\nu}$ and perturbation metric $h_{\mu\nu}$: $g_{\mu\nu} = \eta_{\mu\nu} + h_{\mu\nu}$, $|h_{\mu\nu}| \ll 1$, $T_{00} = -\rho c^{2}$, $T_{ij} = 0$, $\eta_{00} = -1$ \cite{capo15,wu18}, where $c$ is a speed of light. The paper \cite{capo15} reviews the formulation of hybrid metric-Palatini approach and its main achievements in passing the local tests and in applications to astrophysics and cosmology, and in \cite{wu18} the gravitational field equations for the modified gravity $f(R,T)$ theory are considered in the framework of the Palatini formalism.
 
Basic properties of Newtonian limit are: $\phi = \phi_0 + \psi$, \quad $\phi \gg \psi$, \quad $\dfrac{3 \partial_\mu\phi  \partial_\nu\phi}{2\phi} = \dfrac{3 \partial_\mu\psi \partial_\nu\psi}{2\phi} \ll 1$. We denoted the asymptotic of $\phi$ as $\phi_0$, and the local perturbation as $\psi$. Accordingly, Eq. (\ref{equA15}) obtained the following shape of linear order:

\begin{equation} 
- \Box \psi + \left(2V(\phi) - (1 + \phi)V_{\phi} \right) \dfrac{\psi}{3} = \dfrac{\phi_0 \kappa T}{3}
\label{equA16}
\end{equation}

We have neglected the time derivatives of $\psi$, so Eq. (\ref{equA16}) can be written in the following way:

\begin{equation} 
\Delta\psi - m_{\phi}^2 \psi = -\dfrac{\phi_0 \kappa M \delta (r)c^{2}}{3},
\label{equA17}
\end{equation}

\noindent where $m_{\phi}^2 = \dfrac{1}{3} \left(2V(\phi) - (1+\phi)V_{\phi} \right) \big |_{\phi = \phi_0}$ and $T = \rho c^{2} = Mc^{2}\delta(r)$. It can be shown that the effective mass can be expressed in the form: $m_\phi^2 = (2V - V_\phi - \phi(1 + \phi)V_{\phi\phi})\big |_{\phi = \phi_0}$, where $V$, $V_\phi$ and $V_{\phi\phi}$ are the potential and its first and second derivatives with respect to $\phi$, respectively. Solving the equation (\ref{equA17}), it is obtained:

\begin{equation}
\phi = \phi_0 + \psi = \phi_0 + \dfrac{2G\phi_0 M}{3 c^{2}} \dfrac{e^{-m_{\phi}r}}{r}.
\label{equA18}
\end{equation}

\noindent Since the background is Minkowskian, the perturbed Ricci tensor is given by $\delta R_{\mu\nu} = \dfrac{1}{2} (\partial_{\sigma} \partial_{\mu}h_{\nu}^{\sigma} + \partial_{\sigma} \partial_{\nu}h_{\mu}^{\sigma} - \partial_{\mu} \partial_{\nu}h - \Box h_{\mu\nu}) \approx - \dfrac{1}{2} \Delta h_{\mu\nu}$ and $\dfrac{\partial^2 h}{\partial t^2} \approx 0$, $\dfrac{\partial^2 \psi}{\partial t^2} \approx 0$ (slow motion) \cite{wu18,capo15}. Using the following gauge conditions: $\partial_\lambda \widetilde{h}_\mu^\lambda - \dfrac{1}{1 + \phi_0}\partial_\mu \psi = 0$, where $\widetilde{h}_\nu^\lambda \equiv h_\nu^\lambda - \dfrac{1}{2}\delta_\nu^\lambda h_\alpha^\alpha$ \cite{capo15}, Eq. (\ref{equA13}) becomes: 

\begin{equation}
-\dfrac{1}{2}{\Delta}h_{\mu\nu}(1 + {\phi_0}) = \kappa(T_{\mu\nu} - \dfrac{1}{2}{\eta_{\mu\nu}}T) + \dfrac{1}{2}{\eta_{\mu\nu} (V(\phi) + \Delta\phi)},
\label{equA19}
\end{equation}

\noindent from where we obtain:

\begin{equation}
\Delta h_{00} = - \dfrac{2\kappa}{1+\phi_0} (T_{00} - \dfrac{1}{2} \eta_{00}T) + \dfrac{-2\eta_{00}}{2(1+\phi_0)} (V_0 + \Delta \psi),
\label{equA20}
\end{equation}

\noindent where $V_0$ is the minimum of potential $V$ \cite{bork16}, and then

\begin{equation}
h_{00} = -\dfrac{\kappa M c^2}{1+\phi_0} \dfrac{1}{4\pi r} + \dfrac{V_0}{1+\phi_0} \dfrac{r^2}{6 l_c^{2}} + \dfrac{\psi}{1+\phi_0},
\label{equA21}
\end{equation}

\noindent where $l_c$ is a characteristic length scale, corresponding to cosmological background.

\noindent By equating $2\Phi(r)/c^{2} = h_{00}$, we have: 
\begin{eqnarray}
2\Phi(r)/c^{2}&=& -\dfrac{2GM}{1+\phi_0} \dfrac{1}{c^{2}r} + \dfrac{V_0}{1+\phi_0} \dfrac{r^2}{6l_c^{2}} + \dfrac{2 G\phi_0 M}{3 (1+\phi_0) c^{2}} \dfrac{e^{-m_\phi r}}{r} \nonumber \\
&=& -\dfrac{2G_{eff} M}{c^{2}r} + \dfrac{V_0}{1+\phi_0} \dfrac{r^2}{6l_c^{2}},
\label{equA22}
\end{eqnarray}

\noindent with an effective gravitational constant introduced $G_{eff} = \dfrac{G}{1+\phi_0} \left(1- \dfrac{\phi_0}{3} e^{-m_\phi r} \right)$. The term in Eq. (\ref{equA22}) proportional to $r^2$ corresponds to the cosmological background and it could be neglected on galactic level \cite{capo15}.

The modified gravitation potential of Newtonian limit is: 
\begin{equation}
\Phi(r) \approx -\dfrac{G_{eff} M}{r} = - \dfrac{G}{1+\phi_0} \left(1 - \dfrac{\phi_0}{3} e^{-m_\phi r} \right) \dfrac{M}{r}.
\label{equA23}
\end{equation}

\end{paracol}

\reftitle{References}

\end{document}